\documentclass[sigconf]{acmart}

\usepackage{booktabs,threeparttable}

\AtBeginDocument{%
  \providecommand\BibTeX{{%
    \normalfont B\kern-0.5em{\scshape i\kern-0.25em b}\kern-0.8em\TeX}}}

\copyrightyear{2020}
\acmYear{2020}
\setcopyright{none}

\renewcommand\footnotetextcopyrightpermission[1]{} 
\acmConference{X}{X}{X}

\begin{document}

\title{Marketplace for AI Models}

\author{Abhishek Kumar}

\orcid{0000-0003-4383-7225}
\affiliation{%
  \institution{Department of Computer Science \\University of Helsinki}
}
\email{abhishek.kumar@helsinki.fi}

  \author{Benjamin Finley}
\affiliation{%
  \institution{Department of Computer Science \\University of Helsinki}
}
\email{benjamin.finley@helsinki.fi}

\author{Tristan Braud}
\affiliation{%
  \institution{Department of Computer Science \\HKUST}
}
\email{tbraud@cse.ust.hk}

  \author{Sasu Tarkoma}
\affiliation{%
  \institution{Department of Computer Science \\University of Helsinki}
}
\email{sasu.tarkoma@helsinki.fi}

\author{Pan Hui}
\affiliation{%
  \institution{Department of Computer Science \\University of Helsinki \& HKUST}
  }
  \email{pan.hui@helsinki.fi}

\renewcommand{\shortauthors}{Kumar et al.}

\begin{abstract}
Artificial intelligence shows promise for solving many practical societal problems in areas such as healthcare and transportation. However, the current mechanisms for AI model diffusion such as Github code repositories, academic project webpages, and commercial AI marketplaces have some limitations; for example, a lack of monetization methods, model traceability, and model auditabilty. In this work, we sketch guidelines for a new AI diffusion method based on a decentralized online marketplace. We consider the technical, economic, and regulatory aspects of such a marketplace including a discussion of solutions for problems in these areas. Finally, we include a comparative analysis of several current AI marketplaces that are already available or in development. We find that most of these marketplaces are centralized commercial marketplaces with relatively few models.

\end{abstract}




\maketitle
\section{Introduction}

Artificial intelligence (AI) is predicted to have a major societal impact over the coming decades. Specifically, when widely diffused, AI models have the potential to solve many ubiquitous problems in domains ranging from healthcare to transportation. For example, neural network based models have shown human and super-human level performance in many health related diagnostic tasks such as breast cancer detection~\cite{alloghani2020}.

However, to achieve widespread diffusion of AI models and thus capture these benefits, an efficient diffusion mechanism is required. Unfortunately, many popular diffusion mechanisms such as collections of Github repositories, academic research project pages, and existing commercial AI marketplaces have major limitations. Github repositories and academic research projects generally do not have straightforward monetization methods and installing, configuring, applying, and supporting models from these sources can be cumbersome (as such concerns are usually not paramount to academic researchers). While existing commercial AI marketplaces (refer to Section~\ref{sec:existing_marketplaces}) are often centralized and controlled by a single company that may have different motives from the majority of the marketplace users thus allowing for conflict and a single point of failure. Additionally, the models available on commercial AI marketplaces may lag behind the state-of-the-art models available on Github or project pages. Finally, the datasets for training such AI models often cannot be provided for privacy or other reasons, thus the traceability and auditability of models from these sources is lacking. Traits such as these will be important given new AI regulations currently in development by, for example, the European Union (refer to Section~\ref{sec:aimarket_reg}).

\begin{figure}[t]
    \centering
    \includegraphics[width=.5\textwidth]{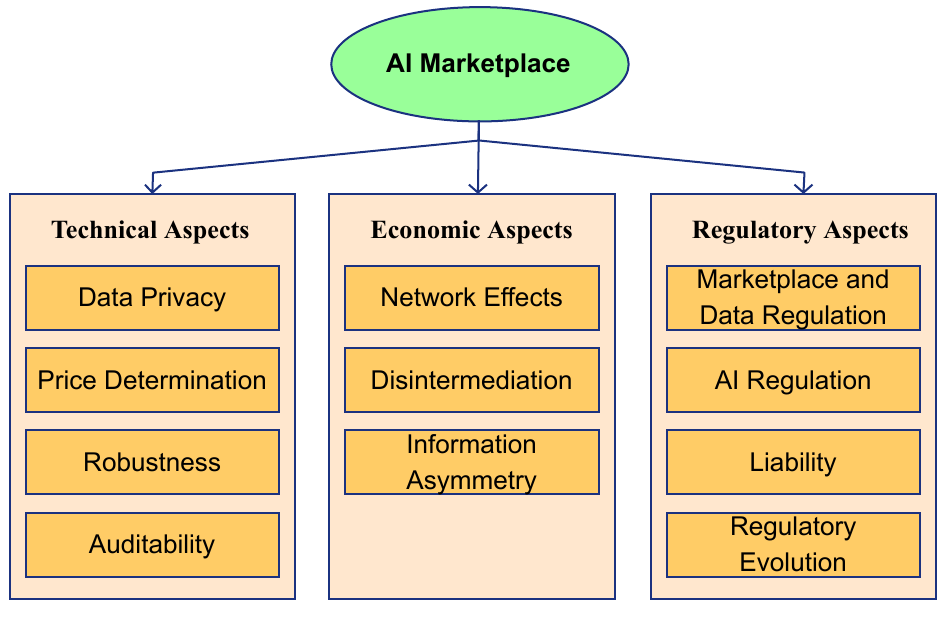}
    \caption{Aspects of an AI Marketplace}
    \label{fig:aspects}
\end{figure}

Given the limitations of existing mechanisms, in this paper we aim to sketch guidelines for a new AI diffusion mechanism based on a decentralized online marketplace and hereafter known as \textbf{AI marketplace}. We consider the technical, economic, and regulatory aspects of creating such a marketplace in order to reach the goal of broad yet ethical AI diffusion (as shown in Figure~\ref{fig:aspects}). The AI marketplace we propose would bring together various actors, including developers and companies of different sizes, users, data collectors, and even governmental entities, towards this common goal.

\section{What is AI Marketplace?}
In our vision, the AI marketplace should facilitate two high level operations: 1) a developer should be able to sell their pre-trained AI models in the marketplace, and 2) a customer should be able to request a custom AI model which suits their specific needs, and the marketplace should be able to match such customers with developers who could build such a model. Also the marketplace should facilitate, when necessary, data sharing (selling and buying) and participation of marketplace actors in, for example, federated learning in support of the mentioned operations.

In a high-level sense, an AI marketplace is similar to other online marketplaces like eBay in terms of business operations and dynamics (e.g., two-sided network effects). However, at the same time, such a marketplace differs from those markets because of the nature of the products. Specifically, an AI marketplace is different from a conventional online marketplace in the following ways:

\begin{itemize}
    \item Developing AI models often requires the sharing of data from the customer side, and such data may be proprietary. Therefore, an AI marketplace may have a mechanism which ensures that developers use that data only for training purposes.
    \item An AI marketplace needs a mechanism which can determine the quality of a final delivered AI model. Conventionally, accuracy has been the primary metric. However, alternative metrics that capture reliability, robustness, and fairness are also now considered important.
    \item Like in any conventional software system, AI systems also require maintenance over time. In the standard software industry, the company which originally developed the software is usually responsible for providing support. However, in an AI marketplace, specific AI developers may not be available in the future. So, an AI marketplace needs standard guidelines which AI developers should follow while developing models for marketplace customers. Thus maintenance by other AI developers is much easier.
\end{itemize}

An AI marketplace could also be considered similar to a mobile app store like Google Playstore and Apple App Store. These stores host many AI-enabled applications, however, they still differ from an AI marketplace in some significant ways:

\begin{itemize}
    \item Unlike in an app store, an AI marketplace could allow customers to request new products on the fly. An AI marketplace can then quickly match AI customers with AI developers with relevant expertise.
    \item Some AI models are proprietary. So, unlike in an app store, they can not be shared online with a wider audience, as this would further facilitate adversarial attacks and risk leaking intellectual property.
\end{itemize}

Finally, an AI marketplace is also similar to an online data marketplace. For example, in both marketplaces the products sold could leak private user data. Specifically, attacks against AI models like model inversion~\cite{fredrikson2015model}\cite{veale2018algorithms} and membership inference~\cite{nasr2019comprehensive}\cite{shokri2017membership} can extract information about users whose data was used in training. However, data marketplaces can still be considered closer to conventional online marketplaces, rather than AI marketplaces, since the transferring of data can be considered analogous to the transferring of physical products (assuming that privacy regulations have been followed) as after the transfer (or successful trade) little maintenance is needed, unlike for AI models.

\section{Driving forces behind AI Marketplace}

An AI Marketplace aims to respond to several issues within the current AI community. Specifically, in order for AI to diffuse and achieve widespread adoption, it is necessary to address the following concerns:

\textbf{Lack of interoperability standards: }Currently, there are multiple frameworks for developing diverse AI models. Different developers use different frameworks (TensorFlow, PyTorch, Caffe2), different languages (Python, Java, C/C++), and target different environments (powerful Linux server, smartphone, minimalist IoT device) depending on the intended usage. Each of these elements comes with challenges. A small AI company may want to pipeline a server-based PyTorch AI model with an externally developed smartphone-based TensorFlow AI model. However, the current lack of interoperability standards dramatically limits such opportunities, and adapting existing models is a tedious task (potentially including redeveloping an application-specific AI model)~\cite{gambo2011lack}\cite{lehne2019digital}.

\textbf{Lack of infrastructure for AI data cooperation: }Many state-of-the-art AI models, especially those based on deep learning, require very large datasets. Unfortunately, the creation, management, and sharing of very large datasets is often difficult for many AI developers (due to resource or capability limitations). As a result, AI development with large data is dominated by researchers in large organizations that have significant capabilities and resources.

\textbf{Rise of data protection regulations: }Collecting user's data is increasingly difficult due to privacy regulations around the world, such as the European General Data Protection Regulation (GDPR) or the California Consumer Privacy Act. These regulations put the burden of protecting the user's privacy on the shoulders of data collectors. Users may give consent to collect their personal data at a given point in time, but they can also withdraw their consent later. The data collector is then often required to erase the collected data. Furthermore, fulfilling these requests requires both technical expertise and regulatory expertise. Most AI developers do not possess both of these skills~\cite{mehri2018privacy}.

\textbf{Large cost of AI development and operation: } According to a Teradata survey, the lack of IT infrastructure (40\%) and the lack of talent (34\%) are the two most significant barriers to AI realization~\cite{teradata2017}. The current lack of qualified AI professionals makes it expensive to hire an AI team. A fully-fledged AI team not only consists of AI developers, but also domain experts, data engineers, product designers, AI sociologists, and IT lawyers. Most small businesses can not afford to hire such an expensive team~\cite{aipricing2020}. Additionally, as mentioned, the infrastructure to collect and store the massive amounts of data required for model development, training, and operation is also costly.

An AI marketplace is a potential solution for overcoming many of these barriers. An AI marketplace can make AI models and datasets accessible to end-users and developers, and give developers a way to monetize their models.

\section{Technical Aspects of AI Marketplace}\label{sec:tech_aspects}
In a basic AI marketplace setting, an AI customer may arrive with a training dataset, and another smaller dataset referred to as a ``validation'' dataset, and want to build a prediction model that performs well on this validation dataset. The AI marketplace can match the customer with AI developers with the skills needed to build a model. If the model developed by the AI developer based on the training dataset (provided by the customer) achieves sufficiently high accuracy (or other metric(s) with a threshold set by the customer) on the validation dataset, then customer and developer can move forward with the transaction.

However, the potential AI model (e.g., a deep neural network like ResNet or Inception) may require a large amount of training data which the customer may not have. Furthermore, the original training dataset provided by the customer may itself be from multiple private sources (e.g., mobile crowdsensing, shopping patterns) and may follow some multimodal distribution. The AI marketplace should also be able to help the customer by allowing the aggregation of multiple alternative datasets from the marketplace while also ensuring the aggregate dataset follows a similar distribution (e.g., using a transfer learning approach~\cite{wang2020transfer}) as the validation dataset. This is important since AI learning algorithms suffer from major model quality loss (or even divergence) when trained on non-IID data~\cite{DBLP:journals/corr/abs-1910-00189}. In the process, the marketplace should also enable the monetization of data in a trusted, fair manner while preserving data ownership and privacy as much as possible~\cite{data2015danger}.

Let us consider the following scenario in the healthcare domain. The consumer is a newly established cancer treatment hospital, and the data sources are cancer institutions from different geographical locations across the globe. The goal of the new hospital is to construct an ML model that can predict the early onset of a given form of cancer. The model must perform well given the demography of its patients, and therefore it is crucial to collect data similar to the small validation set that is representative of the demography. However, the individual data sources have widely different demographics data due to their locations. The goal of an AI Marketplace in such a setting is to enable the collection of a dataset sampled from these sources that matches the demography of the new hospital and in the process attributes fair value to the different data sources. In other settings, the data owner or consumer may not have sufficient AI expertise or skills. The marketplace should be able to connect the consumer to AI experts, and at the same time, should provide a platform for AI experts to assess the performance of their models on the consumer's validation dataset without direct access to the dataset (by the AI experts).

A model/data exchange mechanism in an AI marketplace should have the following properties:

\begin{itemize}
    \item  All individual data sources and the consumer should have privacy protection in the form of differential privacy guarantees to communicate with a specific actor (data owner, AI expert, customer). Such differential privacy guarantees can be achieved by ensuring that all data transferred (between the aggregator and various data owners) is entirely anonymized, de-identified, and ideally encrypted.
    \item Transfer learning should be facilitated in the sense that the aggregating entity acquires a summary training dataset that is statistically related to the consumer validation dataset with respect to the requested metrics.
    \item The aggregator can only learn the pairwise Euclidean distances between the points in the training dataset.
    \item Consumers having sufficient AI skills, but no or little training data should be able to leverage federated learning on the marketplace to facilitate learning without moving raw training data from its original owners. However, to facilitate quick convergence, the marketplace should locate data sources which are nearly identical and independently distributed.
    \item All entities, i.e., AI model contributor or data contributor, should be awarded fairly for their contribution. Unlike normal goods in online marketplaces like eBay, deciding the fair award (or data/model price) is non-trivial in an AI marketplace.
    \item No collaborating entities, i.e., data provider or model training provider in case of federated learning, should be able to cheat in the model building process. In other words, the marketplace should provide robustness against data providers or training update providers who commit malicious or low quality contributions. Thus the marketplace should incorporate a verification mechanism which will assess the quality of the contribution, and hence will determine rewards accordingly.
\end{itemize}

In addition to the properties mentioned above, AI marketplace should also incorporate other relevant properties from conventional marketplaces, e.g., ensuring liquidity in the market~\cite{vulkan2003economics}, providing a framework for conflict resolution between consumer and seller~\cite{tzanetakis2016transparency}. In the remainder of this section, we discuss potential solutions to address these properties.

\subsection{Maintaining data privacy in AI marketplace}
In an AI marketplace context, two major strategies can ensure data privacy for the involved actors. 

\textbf{Federated learning/Peer-to-Peer (P2P) learning as learning paradigms: }Under the federated learning framework~\cite{konevcny2016federated}\cite{hard2018federated} or the P2P learning framework~\cite{bellet2018personalized}\cite{DBLP:journals/corr/abs-1811-11124}, raw data never leaves the owners' devices, and thus data privacy and ownership of the original raw data is always ensured with data owners. However, such paradigms add computational overload in the form of local training on the data owner's end.

\textbf{Using contextual integrity as design principle for data sharing: } The idea of privacy trading is also gaining popularity~\cite{beresford2011mockdroid}\cite{jin2019if}. Many users who are less privacy-sensitive may be willing to sell their privacy. However, even such users should know the future use of their raw data before selling it. The principle of contextual integrity allows enforcing data privacy by providing a framework for evaluating the flow of personal information between different recipients and explaining why certain patterns of information flow are acceptable in one context but problematic in another~\cite{nissenbaum2004privacy}\cite{doi:10.1177/2056305118768300}. The contextual integrity framework allows users to maintain control of their data even after trading it.

\subsection{Price determination in AI marketplace}
Ensuring that the marketplace is transparent in assigning value based on the quality of the data or model relative to the target is extremely important. Unlike in conventional online marketplaces, deciding the right price of data or an AI model is non-trivial. In conventional online marketplaces, the price in the offline market can serve as a reference. Also, in a corporate AI setting, experts can estimate the resources required for a project and hence negotiate the price of AI models with potential buyers. However, in an online AI marketplace, an individual AI developer may not possess the skills to determine the right price for their model.

To ensure fairness when deriving the price, the marketplace should provide a bidding mechanism. This bidding mechanism should ideally 1) be a dominant strategy and incentives-compatible. The dominant strategy of each entity is to bid the amount equal to their private valuation. Bidding this true valuation always leads to a non-zero utility for any entities, 2) maximize the social surplus when all entities report their valuations truthfully, and 3) be implementable in polynomial (preferably linear) time in order to enable scalability. Any mechanism satisfying these three conditions can be said to have employed Vickrey auction~\cite{vickrey1961counterspeculation}. 

Another approach could be to organize contests among AI developers for some pre-defined reward. However, contests and their corresponding reward incentives should be designed based on accurate models of AI developers' strategic behaviour to elicit the desired outcomes~\cite{ghosh2016optimal}. Depending on the strategic behaviour of the AI developers, different kinds of contests can be organized, e.g., contests which reward a fixed number of AI developers, contests which take the form of a tournament, or contests which award everything to the winner~\cite{easley2015behavioral}\cite{gurtler2010optimal}.

\subsection{Robustness against malicious entities in AI marketplace}
A Federated learning or P2P learning framework ensures data privacy and ownership of the data to the original owners. However, such a framework also introduces a number of vulnerabilities. Some entities may want to free-ride by trying to capture benefits (or payments) without making honest contributions, e.g., sending random training updates to the server instead of updates calculated on real data after local training. In other scenarios, other competitors/adversaries may try to introduce model poisoning in order to reduce the reputation of the given marketplace. An AI marketplace should have a verification mechanism to assess whether the given data or training updates are coming from free-riders~\cite{lin2019free}, malicious entities~\cite{bhagoji2019analyzing}, or honest users.

\subsection{Auditability in AI marketplace}
The entire process must be transparent and immutable in order to ensure trust and fairness. Based on how this process is enforced in the marketplace, there exists two types of marketplaces: 1) centralized marketplaces, where a trusted entity ensures smooth operations and maintains an immutable log of all operations on the platform, and 2) decentralized marketplaces, where no single entity is solely trusted by one entity. Instead, all operations are stored on an immutable public distributed ledger (or a public Blockchain). Both marketplace types have their pros and cons. In a centralized marketplace, since the central managing entity earns revenue by charging a small transaction fee for each successful transaction, the entity is motivated to maintain smooth operations on the marketplace by verifying the identity of all parties, improving matching mechanisms, supporting buyer-seller conflict resolution, and ensuring liquidity in the market. In a decentralized AI marketplace, all these goals can be achieved, at the expense of higher energy and time consumption since any changes in the policy must have consensus from peers on the marketplace. Current consensus algorithms are energy-intensive~\cite{CleanTechnica2018}, but these are likely to improve in the future. A decentralized AI marketplace is a true enabler of the vision of democratization of the current tightly concentrated AI ecosystem among a limited number of large players. 

\section{Economic Aspects of AI Marketplace}
Online marketplaces represent an interesting business model in that they facilitate transactions between suppliers and customers often without taking possession of or full responsibility for products or services; thus they have very low cost structures and very high gross margins (e.g., 70\% for eBay, 60\% for Etsy). Additionally, network effects make them highly defensible. For example, Alibaba, Craigslist, eBay, and Rakuten are more than 15 years old, but still dominate their sectors. In the past ten years, the number of online marketplaces worth more than \$1 billion has gone from two (Craigslist and eBay) to more than a dozen in the United States (Airbnb, Etsy, Groupon, GrubHub Seamless, Lending Club, Lyft, Prosper, Thumbtack, Uber, and Upwork). That number is expected to double by the end of 2020, according to Greylock Partners, a Silicon Valley venture capital firm~\cite{hagiu2016network}.

In order to build a successful AI marketplace, a critical number of buyers and suppliers of AI models are needed, just like in any other online marketplace. Potential suppliers for such a marketplace could be AI developers of algorithms and software on platforms like Github, whereas potential buyers could be companies that can not afford their own team of AI experts.

\subsection{Network effects}
A network effect is a phenomenon whereby the value of a platform or service (to an individual participant) is proportional to the number of participants. This phenomenon is often the single-most important factor behind the success or failure of any online platform. Platforms like eBay and Facebook continue to dominate their respective markets partly because they exploit these network effects very well, whereas platforms like Google Plus did not take off partly due to being on the wrong end of such network effects~\cite{zhu2019some}\cite{landeweerd2013success}. So, network effects would be a crucial element for the success or failure of an AI marketplace as well.

The current AI ecosystem favors major players like Google and Facebook as they are already exploiting so called data network effects very well. Specifically, AI-based products or services from these companies become smarter as they train on more data from more users~\cite{mitomo2017data}\cite{li2010network}. In turn, smarter products and services attract more new users, thus creating a feedback loop.

With the adoption of two-tier model training architecture like federated learning~\cite{hard2018federated} (e.g., the first tier being a general global model and the second tier being a personalized local model), challenging the dominance of these companies will be even more difficult. Specifically, this architecture allows these large companies to provide personalized model training to AI customers. The companies provide a pre-trained general model (trained on datasets either owned by the company or procured by the company), and the AI customer personalizes by training on their local dataset. This training paradigm works well even with smaller amounts of a user's training data.

So, if an AI marketplace wishes to challenge the dominance of these companies, it needs to support interoperability between datasets and models so that smaller developers can efficiently aggregate enough smaller training datasets, federated training users, or even models (into an ensemble) to compete (or at least reach a minimum threshold) in terms of performance.

\subsection{Disintermediation}
Conventional online marketplaces fear that once they facilitate a successful transaction, the buyer and the seller will agree to conduct their subsequent interactions outside the marketplace. However, such a risk could be minimized for an AI marketplace. Specifically, many AI models tend to be black box in nature and an AI marketplace could offer buyers independent assessments of the AI model which they are about to procure (thus providing an added-value service). Such assessments might even be mandated by future government regulations as discussed in section~\ref{sec:aimarket_reg}. 

Other strategies could also incentivize AI customers to deal with AI developers through the marketplace. For example, an AI marketplace could also ensure the prices for products without added value (like the mentioned assessments) are the same as when consumers buy directly from the seller or via other channels, otherwise this may also lead to a disintermediation problem.

\subsection{Information asymmetry}
In terms of matching AI customers with AI developers, an AI marketplace is an online freelance marketplace that matches buyers of electronically deliverable services with
freelancers. Just like in any freelancing marketplace, AI customers may also face the issue of ``information asymmetry'', i.e., they may face uncertainty over the quality of individual AI developers. A solution to this dilemma is a trust or reputation mechanism to help facilitate transactions between strangers. Unlike in a conventional online marketplace where transactions mostly involve products/services and monetary payments, transactions in an AI marketplace may also involve training datasets, which may themselves have economic value; therefore mechanisms of an AI marketplace should ensure an even greater degree of trust. Luca et al.~\cite{luca2017designing} found that in an online freelance marketplace, customers are forward-looking and that they place significant weight on a seller's reputation. Though, not controlling for buyers' inter-temporal trade-offs and dynamic selection can considerably bias such reputations. Thus, an AI marketplace should not rely entirely on a reputation mechanism built on reviews from buyers and sellers to tackle the issue of information asymmetry~\cite{yoganarasimhan2013value}. The marketplace needs to play an active role to ensure fairness in these reputations mechanisms. Given that many AI models are black box in nature, the marketplace may need to support independent assessments of an AI model (before final procurement) as suggested in Section~\ref{sec:tech_aspects}, reputation mechanisms could be primarily centered around such assessments in order to ensure fairness.

\section{Regulatory aspects of AI Marketplace}\label{sec:aimarket_reg}
The regulation of AI marketplaces, as well as the regulation of AI in general, is still a significant unknown with major countries only beginning to grapple with the difficult task. The regulation of any AI marketplace as such would combine the regulatory frameworks from several different domains: regulation of the often sensitive training/testing data, regulation of the application of the AI model (often in sensitive domains), and regulation of online marketplaces.

\subsection{Market and data regulation}
The regulation of online marketplaces and sensitive data have historical precedents (e.g., Sherman Act and Health Insurance Portability and Accountability Act (HIPAA) in the USA) due to the analogues of offline marketplaces and physical data. These regulations are thus being overhauled (e.g., GDPR in Europe) to better deal with the issues brought by the internet and AI eras. Therefore, an AI marketplace must comply with these new regulations. As discussed, the fundamental architecture of an AI marketplace can help ensure compliance through data privacy mechanisms such as differential privacy and federated learning and prevention of marketplace monopoly through open interoperability standards (no lock-in and low switching costs). Unfortunately, even current regulations like GDPR are vague in many cases, thus creating uncertainty and cautiousness in companies that may deter participation in an AI market. These uncertainties are compounded by the differing approaches to such regulations in varying markets~\cite{shepardson2020}. We discuss this topic further in Section~\ref{sec:reg_evolution}.

\subsection{AI regulation}
In terms of AI itself, the regulation is currently very sparse but developing. For example, the EU recently released a draft paper outlining it's vision for AI regulation in high-risk areas (e.g., transportation or health care)~\cite{europeancommission2020a}\cite{wallace2020}, as well as ethical guidelines for building trustworthy AI systems~\cite{hleg2019ethics}\cite{kumar2020trustworthy}. The draft includes an overarching framework that would cover the training data (including data traceability and coverage to ensure fairness), model transparency (to understand why certain decisions are taken), and liability (in case of harm).

Similar to the data-centric regulations, an AI marketplace can ensure that at least some subset of the marketplace (e.g., a high-risk area section) enforces or checks that the AI follows these regulations. For example, data traceability can be ensured (at least up to the point of individual marketplace actors) through the use of the public ledger to track the actors providing the data~\cite{singh2018accountability}. Similarly, adequate training data coverage (to prevent discrimination or bias by AI~\cite{ntoutsi2020}) can be ensured through the use of diverse third party (e.g., even government or conformity bodies) validation datasets enforced in a smart contract. In fact, the EU already discusses in the draft paper the potential for ``support structures'' and ``online tools [that] could facilitate compliance'' to help especially small and medium size businesses~\cite{europeancommission2020a}. The marketplace could also periodically re-verify compliance as models evolve as such regulations apply both ex-ante and ex-post. As a final example, data privacy in the visual domain could be ensured by incorporating a mandatory privacy-respecting mechanism in vision-based applications~\cite{10.1145/3204949.3204973}.

Interestingly, in the US, for example, the healthcare domain does have some regulations for AI/ML models partly derived from existing regulations on healthcare software. In fact, many of the models available on existing commercial AI marketplaces are healthcare based (refer to Section~\ref{sec:existing_marketplaces}). As an example of a current regulatory problem, the US regulator is discussing how to regulate AI/ML models that frequently or continuously learn without requiring a regulatory review after every model update (which could be the case under current regulations)~\cite{fda2020}. Again, such future regulations could potentially leverage automated testing on independent government or conformity body validation datasets through an AI marketplace.

Even unconventional organizations are delving into the area, with the Vatican organizing a workshop denoted as ``The `Good' Algorithm? Artificial Intelligence: Ethics, Law, Health''~\cite{mayaki2020}. Additionally, AI regulation and governance has been the subject of recent interdisciplinary academic research by computer scientists, lawyers, and others~\cite{cath2018}\cite{lynskey2019}.

\subsection{Liability regulation}
Liability in the case of such an AI marketplace is also a difficult problem stemming from the difficulty of liability in both AI and online marketplace platforms contexts. Hereafter, given the economic context, we focus on civil liability as opposed to criminal liability. 

Firstly, civil liability, in general, must balance the need to incentivize product safety and compensate victims of harm with the need to encourage business innovation. This balance is especially difficult given the rapid innovation in AI and the potential economic and societal benefits of AI. Additionally, in many legal systems, such as the EU, for compensation the victim must prove damage, a product defect, and a causal link between the two~\cite{europeancommission2020b}. With complex AI or software-based systems, identifying the liable person can be burdensome or in cases with human-AI collaborative systems the liable person may be unclear. A possible solution is to alter the burden of proof requirement, for example, by inverting the burden to rest with the producing company. This inversion would require companies to have very clear and coherent tracking and documenting of AI models which the AI marketplace inherently enables.

In terms of online marketplaces, the liability of companies such as Amazon for products from third-party sellers on their platform (about 58\% of Amazon sales) is a matter of ongoing legal debate~\cite{lehman2020}\cite{busch2019}. For example, in the US, the issue of liability currently revolves around whether Amazon is considered in a legal sense ``a seller'' or simply ``a platform for sellers and buyers". This, in turn, is primarily related to how much power they have over the third party sellers (along with several other considerations). Court cases (e.g., Oberdorf v. Amazon) are currently in progress, and a case may eventually reach the US supreme court. Currently, the status quo in the US is that amazon is not liable. The situation in Europe is similar, with on-going work on developing new regulations and eventually adapting the EU Product Liability Directive~\cite{busch2019}. Given this background, under current trends, if the AI marketplace does not exert strict control and gain excessive power over sellers, then liability could be minimized by maintaining the status of a platform.

\subsection{Regulatory evolution}\label{sec:reg_evolution}
Overall, any AI marketplace would need to evolve along with novel regulation (e.g., safety or export regulations) or risk becoming unusable by legitimate users. Specifically, several distinct marketplaces or strict marketplace access controls may be necessary given new export regulations. For example, new US regulations require companies to have a special license to export certain geospatial AI software~\cite{reuters2020}. The justification for the new regulations is based on national security (with especially China in mind~\cite{medeiros2019changing}). Additionally, a marketplace may need to follow the strictest common safety regulations given the potential for safety regulatory divergence between the US, EU, China, and others. For example, with data privacy, many global internet companies are now GDPR compliant even if they are primarily domiciled elsewhere because they have European interests or customers.

Unfortunately, even in the long term, regulatory convergence may be difficult because AI is also viewed as a strategic security and economic asset to many countries, and thus some do not want to impede any technological progress with regulation~\cite{shepardson2020}.

\section{Where we are now?}\label{sec:existing_marketplaces}
In Table~\ref{tab:existing_marketplaces}, we provide the list of companies which are either providing an AI marketplace or in the process of building such a marketplace. To find such companies, we searched on Google using the keywords: ``online marketplace'', ``data marketplace'', and ``AI marketplace'' and extracted the first ten pages. We then manually visited all links from these pages and checked which describe an entity providing for the trading of AI models, a service to enable trading of AI models, or are in the process of building either of these. After six pages, the search results no longer provided any meaningful links. Eventually, we identified the 24 companies or frameworks listed in Table 1.

Most of these companies are based in either the USA or Europe. Among those which are currently available, none list more than 24 different models (Nauance Communications lists 24, Gravity AI lists 12, and IBM Imaging lists five models), thus suggesting that none have seen major or widespread adoption. The marketplaces which are somewhat mature primarily focus on the healthcare domain. However, these marketplaces are not operating in multiple countries potentially due to the need for regulatory approval of such models in each country or economic area (for example, by the US Food and Drug Administration).

The overarching goal of most of these AI marketplaces aligns with our vision of a general marketplace where buyers and sellers engage in transactions for AI models. In terms of the technical aspects for a successful AI marketplace, from Section~\ref{sec:tech_aspects}, most of the marketplaces have not yet incorporated these though they do often acknowledge the need for such aspects. For example, only two of them mention that they support scalable privacy-preserving model training paradigms like federated learning. As for architecture, most of the marketplaces are proprietary and are based on a centralized architecture. Though the few decentralized marketplaces are primarily based on distributed ledgers, similar to our vision. With regard to the pricing mechanism, most of them are using fixed pricing per model.

\begin{table*}[t]
\centering
\begin{threeparttable}
\caption{Existing AI marketplace frameworks and implementations}
\label{tab:existing_marketplaces}
\begin{tabular}{lllllll}
\toprule
\textbf{AI Marketplace} & \textbf{Domain Focus} & \textbf{Org Type} & \textbf{Architecture} & \textbf{Target Market} & \textbf{Domicile} & \textbf{Status}\\ 
\midrule
Nuance Communications & Diagnostic Imaging & For Profit & Centralized & USA \& Canada & USA & Online\\
Agorai & Various\tnote{a} & For Profit & Centralized & Global & Singapore & Online\\
IBM Imaging &  Healthcare & For Profit & Centralized & USA & USA & Online\\
Envoy AI & Medical Imaging & For Profit &Centralized & USA & USA & Online\\
GraphGrail AI & Various\tnote{b} & For Profit & Centralized & Russia & Russian & Online\\
Algorithmia & General & For Profit & Centralized & USA & USA & Online\\
Neuromation & Various\tnote{d} & For Profit & Centralized & Global & USA & Online\\
Ocean Protocol & General & For Profit & Decentralized & Global & Singapore & Online\\
OVHcloud AI  & General & For Profit & Centralized & Global & France & Beta\\
Orange AI & General & For Profit & Centralized & Global & France & Beta\\
Kynisys & Various\tnote{c} & For Profit & Centralized & Global & UK & Beta\\
Gravity AI & General & For Profit & Centralized & Global & USA & Beta\\
SingularityNET & General & Non-profit & Decentralized & Global & Netherlands & Beta\\
Modzy & General & For Profit & Centralized & USA & USA & Alpha\\
Alphacat & Fintech & For Profit & Centralized & Global & - & Alpha\\
Bonseyes & General & For Profit & Centralized & Europe & Switzerland & In-development\\
Akira AI & General & For Profit & Centralized & Global & India & In-development\\
Genesis AI & General & For Profit & Centralized & Global & USA & In-development\\
AI Global & General & Non-profit & Centralized & Global & USA & In-development\\
Synapse AI & General & For Profit & Decentralized &  Global & USA & In-development\\
TensorTask & General & Non-profit & Decentralized &  Global & USA & In-development\\
Nomidman & General & For-profit & Decentralized & Global & Estonia & In-development\\
OSA Decentralized & Various\tnote{e} & For-profit & Decentralized & Global & BVI\tnote{g} & In-development\\

DaiMoN & General & Non-profit & Decentralized & - & - & PoC\tnote{f}\\

\bottomrule
\end{tabular}
\begin{tablenotes}
\item[a] Finance, Healthcare, Retail, \& Advertising
\item[b] Finance, Travel, Retail, Advertising, \& Consumer Goods
\item[c] Security \& IoT, Oil \& Gas, Robotics, \& Automobile
\item[d] Surveillance, Retail, Medical Imaging, Industrial Robotics, \& Manufacturing
\item[e] Retail, Manufacturing, \& Consumer Goods
\item[f] Proof of Concept (Academic Work)
\item[g] British Virgin Islands
\end{tablenotes}
\end{threeparttable}
\end{table*}

\section{Conclusion}
In this work, we outlined the principles for a marketplace for AI models based on a decentralized online structure. Such a marketplace could help democratize and diffuse AI technology to the greater society, including to smaller actors (like small and medium-size companies). We discussed the technical, economic, and regulatory aspects to consider while designing such a marketplace. We also described (often novel) technologies and solutions that can help address problems in these aspects. For example, utilizing federated learning for privacy-preserving machine learning across marketplace actors. Finally, we studied the current state of various AI marketplaces and provided a comparative analysis of these marketplaces based on properties such as architecture, domain, and status. We found that most of these currently available marketplaces are centralized and company-driven with relatively few models per marketplace. Thus suggesting that AI marketplaces are still in their infancy.



\begin{acks}
This work has been supported by the 5GEAR project funded by the Academy of Finland (Decision No. 319669) and the FIT project funded by the Academy of Finland (Decision No. 325570).
\end{acks}
\bibliographystyle{ACM-Reference-Format}
\bibliography{main}

\end{document}